\documentclass{kluwer}    

\newdisplay{guess}{Conjecture}

\begin{document}                                                                                   
\begin{article}
\begin{opening}         
\title{Colour ratios of pulsation amplitudes in red giants}
\author{J.P. \surname{Huber}, T.R. \surname{Bedding} \&
S.J. \surname{O'Toole}}
\runningauthor{Huber, Bedding \& O'Toole}
\runningtitle{Colour ratios of pulsation amplitudes in red giants}
\institute{School of Physics, University of Sydney 2006, Australia}

\begin{abstract}
We have analysed the three-colour photometry of 34 pulsating red giants
that was published by Percy, Wilson and Henry (2001).  As is common with
semiregulars, many of these stars have long secondary periods (LSPs) of
several years, in addition to the usual shorter-period pulsations.  We
measured pulsation amplitudes in $VRI$ and found that both the $V/R$ and
$V/I$ amplitude ratios in most stars are systematically lower for the LSPs
that for the shorter periods.  These results must be taken into account by
any theory that attempts to explain LSPs.
 
\end{abstract}
\keywords{stars: AGB and post-AGB -- stars: oscillations}

\end{opening}           

\section{Introduction}  

Many semiregular variables have long secondary periods (LSPs) of several
years, in addition to their shorter-period pulsations \cite{Hou63,KSC99}.
The origin of LSPs is unclear, and may be due to binarity
\cite{WAA99,WAW2001} or to strange pulsation modes caused by
convection-pulsation interaction \cite{Woo2000}.  Multi-colour photometry
may shed some light on this puzzle.

\section{Data Analysis}

\inlinecite{PWH2001} have published three-colour ($VRI$) photometry of 34
pulsating red giants covering more than ten years.  These data are publicly
available.  We have calculated amplitude spectra and, for stars showing
more than one period, we measured the amplitudes of the two highest peaks.
This was done in each of the three wavebands, allowing us to calculate
amplitude ratios.

\section{Results}

Figures~1 and~2 show the $V/R$ and $V/I$ amplitude ratios for all stars
with two periods.  In most cases, the longer period is an LSP, with a value
of several hundred days.  The figures show that in a given star, the
amplitude of the LSP generally has less wavelength dependence than does the
shorter period.  This result must be taken into account by any theory that
attempts to explain LSPs.  Note that EG~And is known to be a symbiotic
binary, and the LSP corresponds to the orbital period.  Interestingly, the
amplitude ratios of EG~And are similar to those of other stars, perhaps
lending some weight to the binary model as the explanation of LSPs.

\begin{table} %
\begin{tabular}{lrl}                                        
\hline
{Name}  &  \multicolumn{1}{c}{Spectral Type} &
\multicolumn{1}{c}{Periods (days)}\\
\hline
30 Her (g Her)       & M6 III        & 871\\
4 Ori     & S3.5/1     & 386, 35\\
BC CMi    & M4 III      & 327, 38\\
EG And     & M2e        & 233, 28\\
$\mu$ Gem     & M3 III     & 2547, 26\\
UW Lyn    & M3 IIIab    & 506, 37\\
V1070 Cyg  & M7 III      & 486, 43\\
V642 Her   & M4 III     & 2547, 25\\

\hline
\end{tabular}
\caption[]{Spectral types and periods for a subset of the stars}\label{parset}
\end{table}

\section{Future Study}

Further analysis, perhaps using Period 98, should give more accurate
amplitude ratios.  Comparison with observed velocities may help determine
which stars are members of a binary system \cite{WAW2001,HLJ2002}.

\begin{figure}

\vspace{18cm} 

\includegraphics{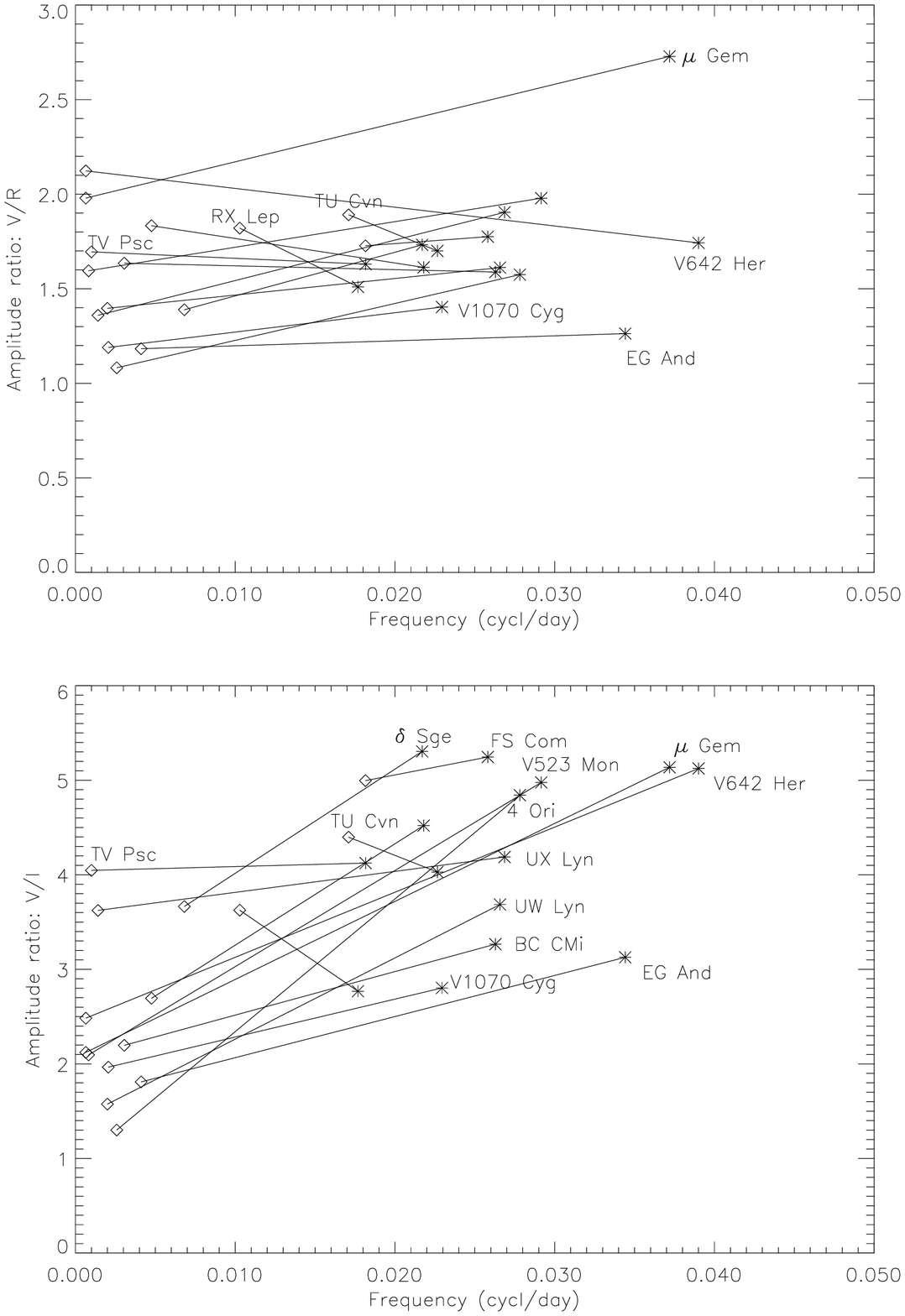}

\caption{Amplitude ratios $V/R$ and $V/I$ for stars with two periods}

\end {figure}

\begin{figure}

\vspace{18cm} 

\includegraphics{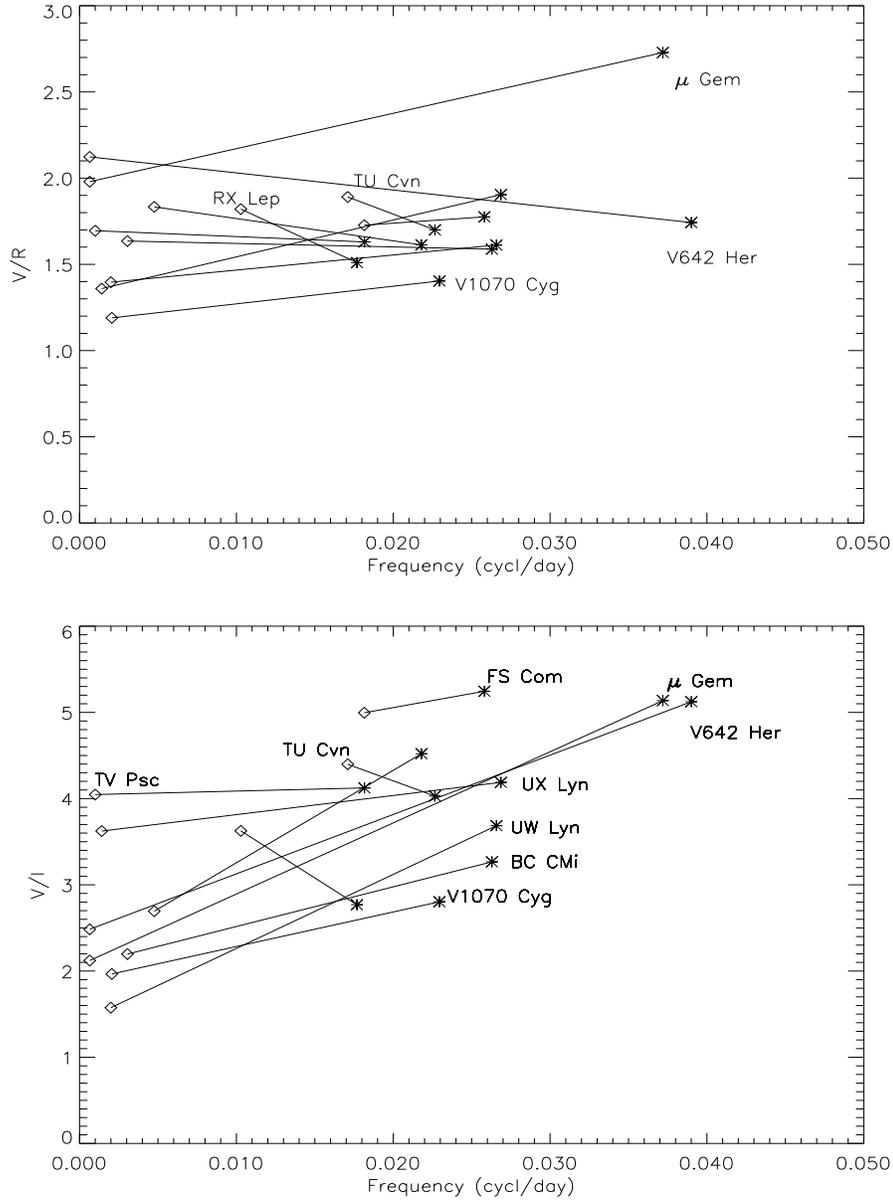}

\caption{Same as Fig.~1, but restricted to stars classified as M giants}

\end {figure}

\end{article}
\end{document}